# Soliton percolation in random optical lattices

Yaroslav V. Kartashov,[1] Victor A. Vysloukh,[2] and Lluis Torner[1]

[1]*ICFO-Institut de Ciencies Fotoniques, and Universitat Politecnica de Catalunya, Mediterranean Technology Park, 08860 Castelldefels (Barcelona), Spain*
[2]*Departamento de Fisica y Matematicas, Universidad de las Americas – Puebla, Santa Catarina Martir, 72820, Puebla, Mexico*
*Yaroslav.Kartashov@icfo.es*

**Abstract:** We introduce soliton percolation phenomena in the nonlinear transport of light packets in suitable optical lattices with random properties. Specifically, we address lattices with a gradient of the refractive index in the transverse plane, featuring stochastic phase or amplitude fluctuations, and we discover the existence of a disorder-induced transition between soliton-insulator and soliton-conductor regimes. The soliton current is found to reach its maximal value at intermediate disorder levels and to drastically decrease in both, almost regular and strongly disordered lattices.

---

Analogies between the electron dynamics in perfect crystals and light propagation in periodic optical media guide the elucidation of a variety of new physical phenomena and related applications [1-4]. Bloch oscillations and Zener tunneling [5-7] are just salient examples of effects that arise in the linear regime of light propagation in periodic optical media, while discrete and lattice solitons [8-13] as well as complex soliton trains [14] are examples of the possibilities afforded by the addition of nonlinearity. All these phenomena occur in regular, periodic or weakly modulated lattices.

Nevertheless, disordered solid-state periodic materials are known to exhibit a wealth of unique electron dynamics phenomena. In optics, the interplay between disorder and nonlinearity was studied, in particular, in systems that may be modeled by the nonlinear Schrödinger equation with random-point impurities [15,16] and in discrete systems [17-19]. Soliton propagation in random potentials and the combined effect of periodic and random potentials on the transmission of moving and formation of stable stationary excitations has been addressed in a number of previous studies (see, for example, Refs. [20-23] and reviews [24-26]). Different regimes of light localization in linear disordered photonic crystals were addressed in Ref. [27]. Band theory of light localization in one-dimensional linear disordered systems was developed in Ref. [28], where it was illustrated that Bragg reflection is responsible for Anderson localization of light. The phenomenon of Anderson-type localization of walking spatial solitons in the nonlinear optical lattices with random frequency modulation was studied in Ref. [29]. Anderson localization in random optical lattices imprinted in photorefractive crystal has been recently observed in a landmark experiment by Segev and co-workers [30]. The phenomena predicted in disordered optical lattices also occur in Bose-Einstein condensates [31-35] and vice versa.

A universal feature of wave packet and particle dynamics in disordered media in different areas of physics is *percolation* [36,37]. Percolation occurs in all types of physical settings, including high-mobility electron systems [38], Josephson-junction arrays [39], two-dimensional GaAs structures near the metal-insulator transition [40], or charge transfer between superconductor and hoping insulator [41], to cite a few.

In this paper we introduce, for the first time to our knowledge, the *nonlinear optical analog* of biased percolation that is related to disorder-induced soliton transport in randomly modulated optical lattices with Kerr-type focusing nonlinearity in the presence of linear variation of the refractive index in the transverse plane, thus generating a constant deflecting force for light beams entering the medium. When such a force is too small, solitons in perfectly periodic lattice are trapped in the vicinity of the launching point due to Peierls-Nabarro potential barriers, provided that the launching angle is smaller than a critical value [42]. Under such conditions soliton transport is suppressed, and thus the lattice acts as a soliton insulator. However, random modulations of the lattice parameters turns soliton transport possible again, with the key parameter determining the soliton current being the standard deviation of phase/amplitude fluctuations. Hereby we discover that the soliton current in lattices with amplitude and phase fluctuations reaches its maximal value at intermediate disorder levels and that it drastically reduces in both, almost regular and strongly disordered lattices. This suggests the possibility of a disorder-induced transition between soliton insulator and soliton conductor lattice states.

Our analysis is based on the nonlinear Schrödinger equation describing propagation of a laser beam in a medium with focusing Kerr-type nonlinearity and spatial modulation of the refractive index in the transverse direction, namely

$$i\frac{\partial q}{\partial \xi} = -\frac{1}{2}\Delta_\perp q - q|q|^2 - Rq - \alpha\eta q. \qquad (1)$$

Here $q$ is the dimensionless complex amplitude of light field; the transverse Laplacian writes $\Delta_\perp = \partial^2/\partial\eta^2$ or $\Delta_\perp = \partial^2/\partial\eta^2 + \partial^2/\partial\zeta^2$ in the case of one- or two-dimensional geometries, respectively; the transverse $\eta,\zeta$ and longitudinal $\xi$ coordinates are scaled in terms of beam radius and diffraction length, respectively; the function $R$ describes the lattice profile, while the parameter $\alpha$ characterizes the rate of linear growth of the refractive index in the direction of $\eta$ axis. Notice that such a refractive index gradient produces a constant deflecting "force" for the light beams entering the medium. Here we concentrate in rather shallow, high-frequency lattices featuring a harmonic regular component with small additive random amplitude or phase fluctuations, with the general functional form $R(\eta) = p\cos(\Omega\eta) + \sigma_a\rho(\eta)$, $R(\eta) = p\cos[\Omega\eta + \sigma_p\rho(\eta)]$, respectively. Here $p$ is the depth of the refractive index modulation, $\Omega$ is the modulation frequency, $\rho(\eta)$ is a random function with zero mean value $\langle\rho(\eta)\rangle = 0$, and unit variance $\langle\rho^2(\eta)\rangle = 1$ (the angular brackets stand for statistical averaging), while the parameters $\sigma_a$ and $\sigma_p$ define the depth of the amplitude or phase stochastic modulations. We assume the correlation function of the random field $\rho(\eta)$ to be Gaussian $\langle\rho(\eta_1)\rho(\eta_2)\rangle = \exp[-(\eta_2-\eta_1)^2/L_{\rm cor}^2]$ with a correlation length $L_{\rm cor}$ larger than the regular lattice period $L_{\rm cor} > 2\pi/\Omega$. In the two-dimensional case, we considered lattices with amplitude and phase fluctuations having the functional shapes $R = p\cos(\Omega\eta)\cos(\Omega\zeta) + \sigma_a\rho(\eta,\zeta)$ and $R = p\cos[\Omega\eta + \sigma_p\rho(\eta,\zeta)]\cos[\Omega\zeta + \sigma_p\rho(\eta,\zeta)]$, respectively, where random field $\rho(\eta,\zeta)$ has the same statistical properties as its one-dimensional counterpart. Such refractive index landscapes can be induced in photorefractive crystals (see Ref. [30]) by combining regular periodic lattices with random nondiffracting patterns. Random component can be generated, e.g., by illuminating a narrow annular slit with a random transmission function placed at the focal plane of a lens. Modification of the transmission function, e.g., by rotating a suitable light diffuser placed after the slit, results in the formation of different random lattice realizations. In the particular case of optical lattices imprinted in SBN crystals biased with a dc electric field $\sim 5\,{\rm kV/cm}$ and laser beam with widths $5\,\mu{\rm m}$ at $\lambda = 532\,{\rm nm}$, a length $\xi \sim 1$ corresponds to some $0.7\,{\rm mm}$, $\Omega = 6$ sets a lattice period $\sim 5\,\mu{\rm m}$, the parameter $p = 1$ corresponds to a refractive index variation $\sim 10^{-4}$, and $q = 1$ corresponds to a peak intensity of the order of $100~\mathrm{mW/cm^2}$.

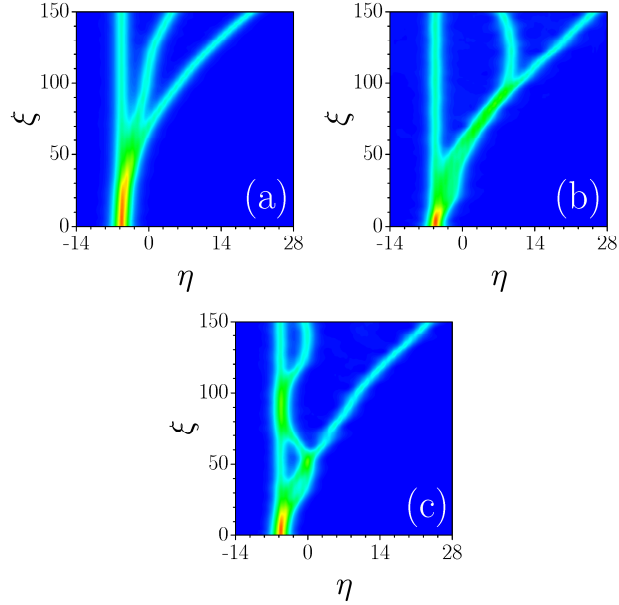

Fig. 1. Soliton intensity distributions corresponding to different realizations of lattices with phase fluctuations at $\sigma_p = 0.03$ (a), $\sigma_p = 0.2$ (b) and amplitude fluctuations at $\sigma_a = 0.05$ (c). In all cases $p = 0.4$, $\alpha = 0.0025$, $\Omega = 6$, and $L_{\rm cor} = 1.8$. Distributions corresponding to different lattice realizations are superimposed.

To gain physical insight into the full nonlinear wave dynamics we start our analysis with an effective particle approach [43], which uses the integral coordinates of the beam center

$$\eta_\mathrm{c} = U^{-1} \int_{-\infty}^{\infty} \int_{-\infty}^{\infty} \eta |q|^2 \, d\eta \, d\zeta, \quad \zeta_\mathrm{c} = U^{-1} \int_{-\infty}^{\infty} \int_{-\infty}^{\infty} \zeta |q|^2 \, d\eta \, d\zeta,$$

where

$$U = \int_{-\infty}^{\infty} \int_{-\infty}^{\infty} |q|^2 \, d\eta \, d\zeta$$

is the total energy flow that remains constant upon propagation. In the deterministic case $\sigma_\mathrm{a,p} \to 0$, the effective particle approach yields coupled equations for the beam center coordinates, which in a two-dimensional geometry write $d^2\eta_\mathrm{c}/d\xi^2 + W\Omega \cos(\Omega\zeta_\mathrm{c})\sin(\Omega\eta_\mathrm{c}) = \alpha$ and $d^2\zeta_\mathrm{c}/d\xi^2 + W\Omega \cos(\Omega\eta_\mathrm{c})\sin(\Omega\zeta_\mathrm{c}) = 0$. Here we take a trial function $|q(\eta,\zeta,\xi)| = q_0 \times \exp[-\chi^2(\eta-\eta_\mathrm{c})^2]\exp[-\chi^2(\zeta-\zeta_\mathrm{c})^2]$, with the amplitude $q_0$ and form-factor $\chi$. The parameter $W = p\exp(-\Omega^2/2\chi^2)$ characterizes the height of the Peierls-Nabarro potential barrier, which grows linearly with the lattice modulation depth $p$ and diminishes rapidly with increasing carrying frequency $\Omega$. Other trial functions provide qualitatively similar results. The evolution equation for the beam center in one-dimensional geometries is obtained by setting $\zeta_\mathrm{c} \equiv 0$. The critical value of rate $\alpha$ of linear increase of refractive index is directly related to height of the barrier, and is given by $\alpha_\mathrm{cr} = W\Omega$. At $\alpha > \alpha_\mathrm{cr}$ the beam starts to drift across the lattice in the positive direction of $\eta$ axis, while at $\alpha < \alpha_\mathrm{cr}$ it is trapped in the vicinity of the launching point. However, in the presence of fluctuations the height of potential barrier becomes a random function of the transverse coordinates, thereby generating a non-vanishing *probability of soliton transport,* or percolation, even at $\alpha < \alpha_\mathrm{cr}$. This phenomenon may be viewed as a disorder-induced transition between soliton-insulator and soliton-conductor regimes. Series of comprehensive numerical simulations for different regions of the parameter space revealed that this transition appears most clearly in shallow, high-frequency lattices, where the soliton mobility is relatively high and the radiative losses are small in unperturbed lattices.

To elucidate the exact implications of this result, we conducted a comprehensive Monte-Carlo numerical investigation, by integrating Eq. (1) numerically with a split-step Fourier method. We propagated solitons up to a given large distance, termed $\xi_\mathrm{end}$, for different series of random realizations of the refractive index profiles $R_k(\eta,\zeta)$, $1 \leq k \leq N_\mathrm{r}$, with $N_\mathrm{r} = 10^3$. We first focus on a detailed quantitative analysis of one-dimensional geometries. We set a lattice frequency at $\Omega = 6$ and launching point at $\eta_0 = -10\pi/\Omega$ (coinciding with one of local lattice maxima). We used a wide integration domain $\eta \in [-100,100]$ to eliminate any boundary effects. An input beam with the shape $q|_{\xi=0} = \mathrm{sech}(\eta - \eta_0)$ and an input energy flow $U = 2$ was selected. Such input would propagate undistorted in a uniform cubic medium. The critical refractive index slope for such beam in the optical lattice of depth $p = 0.4$ amounts to $\alpha_\mathrm{cr} = 0.0027$, and such beam would be trapped in regular lattice with $\alpha = 0.0025$. Notice that the shapes of truly stationary solitons supported by the lattice with a refractive index gradient may be slightly asymmetric and that they are fairly similar to those existing in lattices with a linear amplitude modulation [see, e.g., Fig. 1(d) of Ref. [46]]. Such stationary states exist above a power threshold that depends on the refractive index gradient. For the seek of generality and experimental relevance, in what follows we consider symmetric inputs with a bell-shape in the form $q|_{\xi=0} = \mathrm{sech}(\eta - \eta_0)$.

Figure 1 shows an illustrative set of soliton intensity distributions in a lattice with random phase and amplitude fluctuations for different standard deviation values obtained by direct numerical integration of Eq. (1). When the local restoring force acting on the soliton in the random lattice compensates the random-induced deflecting force, the soliton remains trapped in the input lattice channel. Otherwise, it is accelerated and moves along a complex path. The kinetic energy of the associated effective particle gradually increases and the probability of its trapping in one of subsequent lattice channels or backward reflection in the vicinity of regions with locally decreased refractive index decreases. Yet, it does not vanish completely [see, e.g., Fig. 1(b)]. The radiation, that appears unavoidably when solitons cross the lattice channels, is weak until the local propagation angle does not approach the Bragg angle $\theta_\mathrm{B} = \Omega/2$. Note that the soliton propagation trajectories approach parabolic ones in lattices with weak fluctuations $\sigma_{\mathrm{a,p}}$ and that they can be quite complex already when $\sigma_{\mathrm{a,p}} \sim 0.2$. In the case of quasi-parabolic trajectory the overall displacement of soliton can reach values $\sim \alpha\xi^2/2$, which for a propagation distance $\xi \sim 150$ can exceed the input soliton width by more than one order of magnitude, as readily visible in Fig. 1(a).

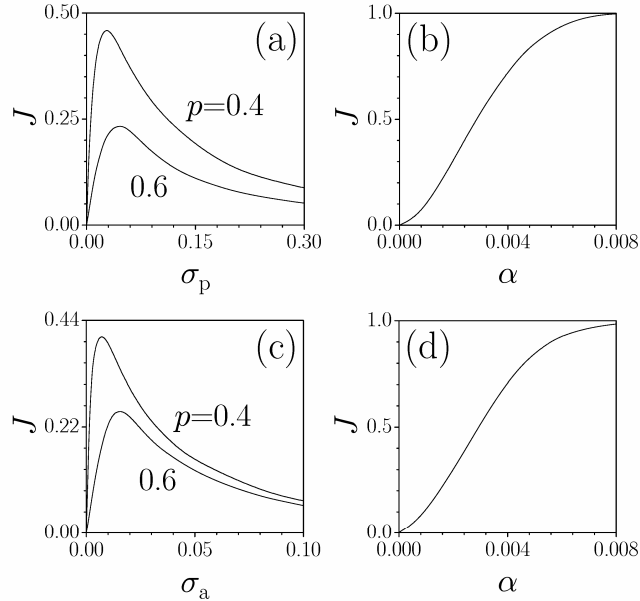

Fig. 2. Soliton current versus dispersion of phase fluctuations at $\alpha = 0.0025$ and various lattice depths (a) and versus slope of linear potential at $\sigma_\mathrm{p} = 0.03$ and $p = 0.4$ (b). Soliton current versus dispersion of amplitude fluctuations at $\alpha = 0.0025$ and various lattice depths (c) and versus slope of linear potential at $\sigma_\mathrm{a} = 0.007$ and $p = 0.4$ (d). In all cases $\Omega = 6$, $L_\mathrm{cor} = 1.8$, and soliton current is calculated for $10^3$ lattice realizations.

It is worth stressing that the dynamics depicted in Fig. 1 cannot be considered as an initial stage of Bloch oscillations that take place in linear systems [5-7]. In our model the strongly nonlinear excitations experience only small modulations due to the presence of a shallow lattice and their diffraction spreading is balanced mostly by nonlinearity. Importantly, the corresponding localized excitations move across the regular lattice over the significant distances without changing their shape, provided that their velocities are sufficient to overcome the so-called Peierls-Nabarro potential barrier. If such nonlinear wave packet is accelerated by a constant force and the deflection angle approaches the Bragg one, it spreads rapidly due to strong radiative losses, as shown in Refs. [44,45] instead of being reflected as a single object, as it occurs upon Bloch oscillations in arrays of evanescently-coupled waveguides described by discrete model. Note that when the propagation angle is smaller than the Bragg angle, solitons always preserve their internal structure even upon reflection on the lattice defects [an effect visible in Fig. 1(b)], in contrast to the phenomena generated by Bloch oscillations.

A key percolation feature to be monitored is the dependence of the soliton current on the level of randomness present in the system. We calculated the total number of solitons $N_c$ launched with zero velocity at $\eta_0 = -10\pi/\Omega$, that reach the point $|\eta_0| = 10\pi/\Omega$ at any random distance $\xi_c$ for a set of $N_r = 10^3$ realizations of optical lattices with amplitude or phase fluctuations. This enables to introduce soliton current, defined as $J = N_c/N_r$. Figure 2(a) shows the dependence of such quantity on the standard deviation of phase fluctuations $\sigma_p$ for different values of the lattice depth $p$. Since the value of deflecting force was below the critical value $(\alpha < \alpha_{cr})$, a rapid growth of the soliton current with increasing $\sigma_p$ occurs. This is entirely due to the formation of continuous cluster of opened lattice sites connecting the input and output points. This phenomenon is intuitively analogous to the disorder-induced phase transition that arises between insulating and conducting states. However, further increasing $\sigma_p$ results in a decrease of soliton current due to growing probability of soliton trapping or reflection in the vicinity of lattice sites that are sufficiently strongly distorted.

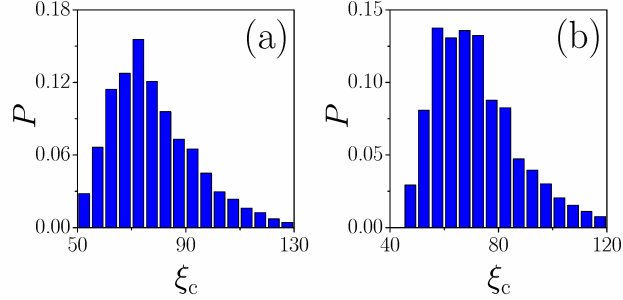

Fig. 3. Statistics of the distance $\xi_c$ for lattices with phase fluctuations at $\sigma_p = 0.1$ (a) and amplitude fluctuations at $\sigma_a = 0.03$ (b). In both cases $\alpha = 0.0025$, $\Omega = 6$, $p = 0.4$, and $L_{cor} = 1.8$.

Therefore, one of the central results of this work is that soliton current in the lattices with fluctuations reaches its maximal value at *intermediate* disorder level and substantially decreases in almost regular $(\sigma_p \to 0)$ and strongly disordered $(\sigma_p \to 1)$ lattices. Growth of the lattice depth $p$ results in a drop off of the maximal value of the soliton current due to the reduction of soliton mobility in deeper lattices. The soliton current maximum is achieved at slightly larger disorder levels in deeper optical lattices. Figure 2(b) illustrates the dependence of the soliton current on the rate $\alpha$ of the linear refractive index variation (i.e., the equivalent to the applied voltage in electro-dynamics). Instead of the sharp jump from zero to unity at $\alpha = \alpha_{cr}$ that takes place in regular lattices, the current is growing smoothly in disordered lattices. Remarkably, the current may be nonzero even for very small slopes $\alpha$ that are far below the critical value and can be less than unity for $\alpha > \alpha_{cr}$. Growth of the disorder level results also in an increase of the transition region of the $J(\alpha)$ dependence that can be defined as a region where the current increases from $0.1$ to $0.9$. We have found qualitatively similar dependencies of the current on the disorder level and slope $\alpha$ in lattices with amplitude fluctuations [Figs. 2(c) and 2(d)]. Notice, however, a remarkably higher sensitivity of the soliton current on the standard deviation in the case of amplitude fluctuations. Note that the results depicted in Fig. 2 remain almost unchanged for much longer integration distances, exceeding $\xi = 150$. This is so because solitons that become trapped at a certain lattice realization typically remain in the vicinity of the launching point for distances that far exceed 150 diffraction lengths, while the probability of trapping for solitons that start walking and being accelerated typically decreases with the propagation distance, so that such solitons will contribute to soliton current on larger distances too.

As mentioned above, the distance $\xi_c$ at which soliton reaches the point $\eta = |\eta_0|$, is a random value because soliton follows a random path to reach it. Figure 3 shows histograms of this random value for lattices with either phase or amplitude fluctuations. The histograms show ratio between the number of lattice realizations where $\xi_c$ falls into fixed intervals, and the total number of events when soliton eventually reaches the point $|\eta_0| = 10\pi/\Omega$. The maximum of the histogram corresponds to the most probable distance at which soliton reaches the point $\eta = |\eta_0|$ (one has $\langle \xi_c \rangle \approx 78$ in the case of phase fluctuations and $\langle \xi_c \rangle \approx 72$ in the case of amplitude fluctuations for selected disorder levels), while the histogram width characterizes the standard deviation of $\xi_c$. We found that in the regime corresponding to a growth of the soliton current with $\sigma_{a,p}$ [see Figs. 2(a) and 2(c)] an increase of the fluctuation level causes increase of the histogram width (standard deviation of $\xi_c$) and an increase of the distance $\langle \xi_c \rangle$ mainly due the expansion of the histogram tail toward the region of higher $\xi_c$ values. Note the analogy of this phenomenon with the scattering of ultra-short pulses in turbid media, where an increase of the turbidity results in larger group delays and pulse broadening. Note also that the sensitivity of $\xi_c$ on the standard deviation is remarkably higher in the case of amplitude fluctuations.

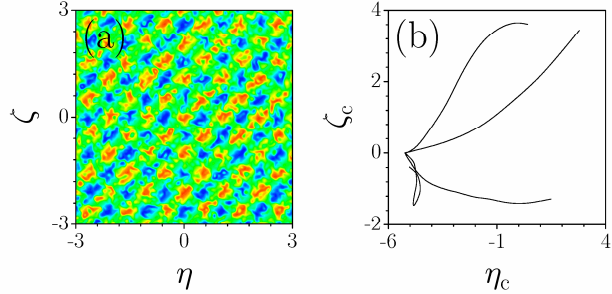

Fig. 4. (a) Lattice with random phase fluctuations at $\Omega = 6$, $\sigma_p = 0.1$, $L_{cor} = 1.8$. (b) Soliton propagation trajectories on $(\eta_c, \zeta_c)$ plane corresponding to different realizations of lattice with phase fluctuations at $\sigma_p = 0.2$, $p = 0.4$, $\alpha = 0.0015$, $\Omega = 6$, and $L_{cor} = 1.8$.

We also investigated the possibility of soliton percolations in two-dimensional geometries. In these geometries, soliton mobility in a medium with cubic nonlinearity is strongly restricted because of the possibility of development of collapse or decay of the input beams. Thus, soliton moving across two-dimensional lattices may drastically broaden in between sites of the lattice and can hardly jump between more than two sites. To avoid such type of behavior, we added phenomenological nonlinearity saturation into Eq. (1) by rewriting the nonlinear term in the form $-q|q|^2/(1+S|q|^2)$. Notice that nonlinearity saturation is typical for photorefractive crystals, where disordered lattices might be imprinted optically. It is worth mentioning that in the previous cases we addressed the canonical model of Kerr nonlinearity in a (1+1)-dimensional geometry because of its generality. The model is directly applicable to Bose-Einstein condensates and to light propagation in photorefractive crystals (with rather intensive background illumination), as well as in suitable semiconductor lattices. In (2+1)-dimensions Kerr models ex-hibit collapse, which may be arrested by saturation, e.g., in photorefractive media with a suitable background illumination.

Figure 4(a) shows a typical profile of a two-dimensional lattice with random phase modulation. The fundamental soliton corresponding to the saturation parameter $S = 0.2$ and energy flow $U = 8$ was launched in the point $\eta_0 = -10\pi/\Omega$, $\zeta = 0$ into such lattice in the presence of linear refractive index variation with slope $\alpha = 0.0015$. Figure 4(b) shows an illustrative set of random trajectories corresponding to different lattice realizations. One of them (a small almost closed loop) illustrates the possibility of soliton trapping in the vicinity of the launching point, while other trajectories confirm that soliton percolation (namely, drift in the positive direction of $\eta$ axis) does occur in two-dimensional settings. One of the characteristic features of two-dimensional soliton percolation, in comparison with one-dimensional case, is a faster increase of soliton current with disorder level due to much richer set of possible propagation paths in the two-dimensional geometry.

We thus conclude by stressing that we have introduced the phenomenon of percolation of spatial solitons in optical lattices with random phase or amplitude fluctuations. The central discovery reported is the possibility of disorder-induced transition between soliton insulator and soliton conductor lattice states. Such phenomenon opens up further analogies with solid-state and statistical physics, whose experimental exploration has become feasible after the recent observation of Anderson localization in random optically induced lattices [30].

**Acknowledgements**

This work has been supported in part by the Government of Spain through the Ramon-y-Cajal program and the grant TEC2005-07815, and by CONACYT through the grant 46552.